\documentclass[prl,twocolumn,superscriptaddress]{revtex4-2}
\usepackage{bbm}
\usepackage{amsfonts}
\usepackage{amssymb}
\usepackage{amsmath}
\usepackage{graphicx}
\usepackage{dcolumn}
\usepackage{appendix}
\usepackage{bm}
\usepackage{units}
\usepackage{txfonts}
\usepackage{multirow}
\usepackage{CJKutf8}
\usepackage{subfigure}
\usepackage{color}
\usepackage{soul}
\usepackage{ulem}
\usepackage{url}
\usepackage[colorlinks,linkcolor=blue,anchorcolor=blue,citecolor=blue,urlcolor=blue]{hyperref}
\usepackage{array}
\begin{document}
\title{On field-driven domain wall motion in compensated ferrimagnetic nanowires}

\author{K. Y. Jing}
\affiliation{Physics Department, The Hong Kong University of Science and 
Technology, Clear Water Bay, Kowloon, Hong Kong}

\author{X. Gong}
\affiliation{Physics Department, The Hong Kong University of Science and 
Technology, Clear Water Bay, Kowloon, Hong Kong}

\author{X. R. Wang}
\email[Corresponding author: ]{phxwan@ust.hk}
\affiliation{Physics Department, The Hong Kong University of Science and 
Technology, Clear Water Bay, Kowloon, Hong Kong}
\affiliation{HKUST Shenzhen Research Institute, Shenzhen 518057, China}
\date{\today}

\begin{abstract}
The fascinating high-speed field-driven domain wall (DW) motion along 
ferrimagnetic nanowires near the angular momentum compensation 
point (AMCP) is solved based on the generic ferrimagnetic dynamics.   
The physics of the absences of precessional torque and infinite high 
Walker breakdown field at the AMCP is proved under general conditions. 
Based on the energy conservation principle, an almost exact DW velocity formula, 
valid beyond the Walker breakdown field, is obtained. Our results agree  
with all existing experiments and simulations. This theory provides 
useful guidances to DW manipulation. 
\end{abstract}

\maketitle

\textit{Introduction}.---
Magnetic domain wall (DW) dynamics in nanowires have attracted much 
attention for its rich physics \cite{hubin,xiansi} and promising device 
applications such as racetrack memories \cite{Parkin1}. One critical 
issue in applications is the realization of high stable DW speed under 
external forces such as magnetic fields and electrical currents. 
This requires a delay or removal of so-called Walker breakdown \cite{Walker}. 
The endeavour of increasing DW speed leads to studying DW motion in 
antiferromagnetic nanowires \cite{AFMReview,AFMDW1,AFMDWexperiment}, and, 
very recently, to that in ferrimagnetic nanowires \cite{Review1,exper1,exper2,
exper3,exper4,exper5,exper6,exper7,Ferrisc,FerriAC,FerriSW,Ferrianisotropy}. 
A ferrimagnet has at least two spin sublattices antiferromagnetically 
interacting with each other. It has two special states called the angular 
momentum compensation point (AMCP) at which the angular momenta of the two 
sublattices cancel each other and the magnetization compensation point 
at which the magnetizations cancel each other. One class of ferrimagnets is 
rare-earth-transition-metal alloys whose AMCP and magnetization compensation 
point are different in general and can be tuned by compositions, other than the 
temperature. Unlike an antiferromagnet, ferrimagnetic states can be manipulated 
by a magnetic field, a spin transfer torque, and a spin-orbit torque. 
Also, unlike a ferromagnet, the net magnetization of a ferrimagnet can be very 
small but not zero, especially around an AMCP such that it is susceptible 
to the magnetic field with small Zeeman energy. One fascinating discovery 
is the very high DW speed of thousands meters per second in compensated 
ferrimagnetic (FiM) nanowires near the AMCP \cite{exper1,exper2,exper3}.
Here we show that high DW speed near the AMCP is related to the absence 
of precessional torque and Walker breakdown phenomenon at the AMCP. 

Although FiM dynamics should be described by coupled partial differential equations 
for magnetizations on at least two antiferromagnetically coupled sublattices, 
existing theoretical studies treat a ferrimagnet either as a ferromagnet 
whose dynamics follows Landau-Lifshitz-Gilbert (LLG) equation \cite{exper2,exper3} 
or an antiferromagnet with the N\'{e}el order governed by a second-order partial 
differential equation \cite{exper1,exper4,Eduardo1,effLLG4,FiMDW1,Ferrisc}. 
DW dynamics is then obtained from converting the partial differential equations into 
ordinary differential equations for the collective coordinates of DW center and 
DW-plane canting angle \cite{exper1,exper4,Eduardo1,effLLG4,FiMDW1,Ferrisc}. 
Indeed, existing theories have enriched our understanding of DW dynamics in 
ferrimagnets in many aspects. However, there are some drawbacks in these approaches. 
These approaches fail to provide a quantitative explanation to both experiments 
and simulations since they rely on the existence of a DW plane and a rigid body assumption for 
the Thiele equation \cite{Thiele}. It often needs to assume also certain DW 
structure such that the approaches are difficult, if not impossible, to generalize 
to situations where the assumptions are not valid such as for vortex DWs and DWs in 
chiral magnets. Furthermore, the physical picture behind the FiM DW motion is 
unclear in these approaches and an accurate description of the DW speed beyond 
the Walker breakdown field is still challenging. 

In this work, the origin of the high DW speed and absence of Walker breakdown 
field at the AMCP of a FiM nanowire are explained based on generic dynamics 
for coupled sublattice magnetizations of a ferrimagnet with a general Rayleigh 
dissipation. We show that a static DW between two domains with different energy 
densities does not exist. Spins in the DW must move in a field that creates such 
an energy density difference. Moving spins must dissipate energy due to the 
inevitable coupling between spins and its environment described by Gilbert damping 
in magnetization dynamics. The dissipated energy must be compensated by the Zeeman 
energy released from the DW propagation toward domain of the higher energy density. 
At the AMCP, precessional torque vanishes due to the zero angular momentum 
and the Walker breakdown field become infinity, leading to the high DW speed. 
Furthermore, a universal relationship between DW speed and DW structure is 
obtained, and an almost exact formula for high-field DW velocity is derived.

\begin{figure}
	\centering
	\includegraphics[width=8.5cm]{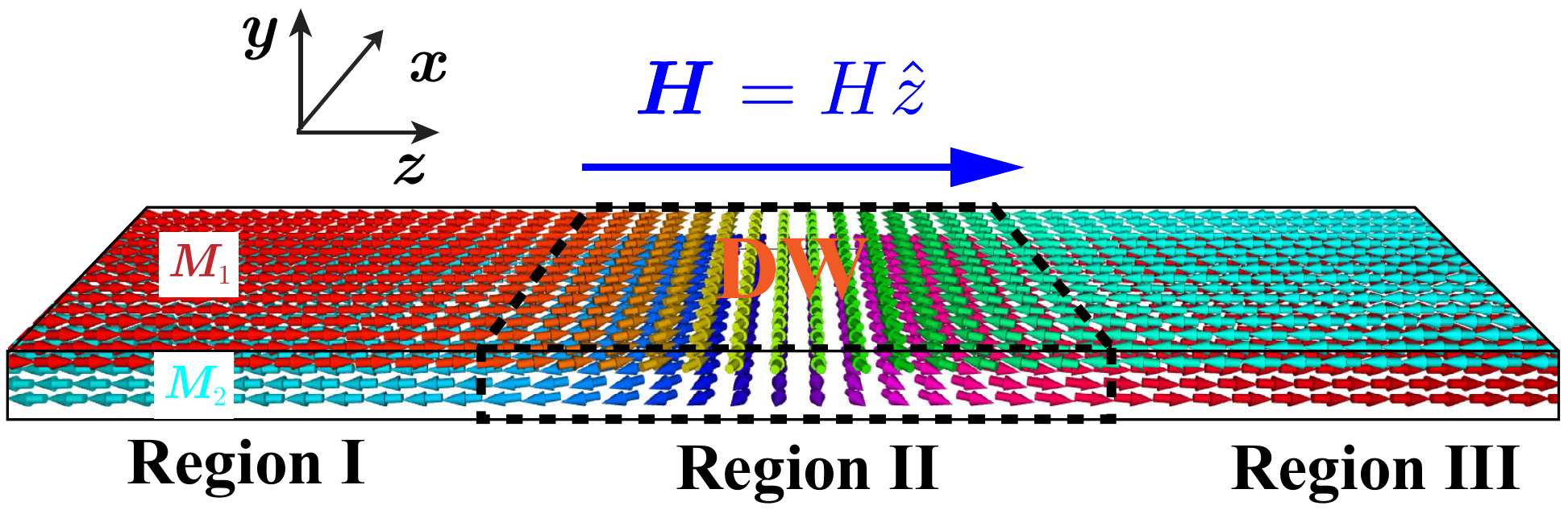}\\
\caption{Schematic of a HH FiM DW in a nanowire. Region I and III are two uniform FiM
domains, separated by a DW (region II) whose width is $\Delta$. DW structure can be very 
complicated. $\boldsymbol{H}$ is the external field. Colours denote the spin orientations: 
The red for spins along $\hat z$ and the light-blue for spins along $-\hat z$. }
	\label{fig1}
\end{figure}

\textit{Model}.---
We consider a head-to-head (HH) DW in a FiM nanowire, whose easy axis is along 
the wire defined as the $z$-axis as shown in Fig. \ref{fig1}. $\boldsymbol{M}_1$ 
and $\boldsymbol{M}_2$ are the magnetizations on two sublattices with $M_1$ and 
$M_2$ being their saturation magnetization. 
The total magnetic energy of the wire in the presence of a uniform magnetic field 
$\boldsymbol{H}$ is $E=\int\varepsilon\, d^3\boldsymbol{x}$ with the energy density of 
\begin{equation}
\varepsilon= J\boldsymbol{M}_1 \cdot\boldsymbol{M}_2 +\sum_{i=1,2}\left[ A_i\left( 
\nabla\boldsymbol{M}_{i}\right)^2 +f_i(\boldsymbol{M}_i)- \mu_0 \boldsymbol{M}_i
\cdot  \boldsymbol{H}\right],
\end{equation} 
where $J>0$ is the antiferromagnetic interlattice-spin coupling constant. 
$A_i$ and $f_i$ are the ferromagnetic exchange stiffness and anisotropic magnetic 
energy density for sublattice $i$ ($i=1,2$). $f_i$ is assumed to have two equal 
minima at $\boldsymbol{M}_i=\pm M_i\hat{z}$.

The FiM magnetization dynamics is generically governed by the following equations 
\cite{Yuan1,AFM1}
 \begin{equation}\label{dynamics}
 \begin{aligned}
\frac{1}{\gamma_1} \frac{\partial \boldsymbol{M}_1 }{\partial t}=- \boldsymbol{M}_1 \times 
\left(  \boldsymbol{H}_1 - \frac{\alpha_{11}}{\gamma_1 M_1 } \frac{\partial \boldsymbol{M}_1 }{\partial t} 
   - \frac{\alpha_{12}}{\gamma_1 M_1 } \frac{\partial \boldsymbol{M}_2 }{\partial t} \right)\\
\frac{1}{\gamma_2} \frac{\partial \boldsymbol{M}_2 }{\partial t}=- \boldsymbol{M}_2 \times \left( 
 \boldsymbol{H}_2 - \frac{\alpha_{22}}{\gamma_2 M_2 } \frac{\partial \boldsymbol{M}_2 }{\partial t}  
   - \frac{\alpha_{21}}{\gamma_2 M_2 } \frac{\partial \boldsymbol{M}_1 }{\partial t}   \right),
\end{aligned}
\end{equation}
where $\boldsymbol{H}_i =-\mu_0^{-1} \delta E/ \delta  \boldsymbol{M}_i$ and $\gamma_i
= g_i \mu_B /\hbar$ ($i=1,2$) are the effective field and the gyromagnetic ratio for 
$\boldsymbol{M}_i$, respectively. $g_i$, $\mu_B$, and $\hbar$ are the Land\'{e} 
$g$-factor of sublattice $i$ ($i=1,2$), the Bohr magneton, and the Planck constant, 
respectively. $\alpha_{11}, \alpha_{22}$ and $\alpha_{12}, \alpha_{21}$ are 
intra-sublattice and inter-sublattice damping coefficients. We have $\frac{\alpha_
{12}}{\gamma_1M_1}=\frac{\alpha_{21}}{\gamma_2M_2}$ due to the action-reaction law. 
$s_i = M_i /\gamma_i$ is the spin density of sublattice $i$ ($i=1,2$).
$\gamma_1 \neq \gamma_2$ in a general ferrimagnet because of the difference in 
Land\'{e} $g$-factors of sublattices. For example, in GdFeCo alloys, 
$g_{\mathrm{Gd}}\simeq2$, $g_{\mathrm{FeCo}}\simeq2.2$ \cite{exper1}. 

\textit{Results}.---
We prove first that no static DW is allowed in the presence of a magnetic 
field along the $z-$direction, except at magnetization compensation point.
If a static DW solution exists, the DW structure should satisfy equations 
$\boldsymbol{M}_i \times \boldsymbol{H}_i =0$ ($i=1,2$).
As illustrated in the Supplemental Materials \cite{supp}, it implies 
$\boldsymbol{M}_i(\boldsymbol{x},t)$ ($i=1,2$) satisfying following equation
\begin{equation}\label{eq_Evar}
\oiint_{\partial \Omega} \left[ \varepsilon \mathbbm{1} - \sum_{i=1,2}^{j=x,y,z} 2A_i 
(\nabla M_{i,j})\otimes (\nabla M_{i,j}) \right] \cdot d\boldsymbol{\sigma} =\mathrm{const.} 
\end{equation}
where $\partial \Omega$ is any closed surface of the system, 
$\mathbbm{1}$ is the $3\times 3$ unit matrix, and $\otimes$ denotes the dyadic product.
Eq. \eqref{eq_Evar} cannot be true for a DW with $\boldsymbol{M}_1=M_1\hat z, \boldsymbol
{M}_2=-M_2\hat z$ on its left and $\boldsymbol{M}_1=-M_1\hat z,\boldsymbol{M}_2=M_2\hat z$ 
on its right as shown in Fig. \ref{fig1}, or vice versa, because it requires 
$(M_1 - M_2)H =0$. Thus, a static DW can only exist either with $H=0$ or $M_1 =M_2$. 
In other words, a static DW cannot exist between two domains with different energy density. 
This result can also be understood from following argument: Assume $\boldsymbol{M}_i(\boldsymbol
{x})$ is a static DW that separate a left domain with a lower energy density $\varepsilon_1$ 
from the right domain with a higher energy density $\varepsilon_2 (>\varepsilon_1)$. 
The energy change by shifting DW to the right by a distance $L$, i.e. $\boldsymbol{M}_i
(\boldsymbol{x})\rightarrow \boldsymbol{M}_i(\boldsymbol{x}+L\hat z)$, 
is $LS(\varepsilon_1-\varepsilon_2)<0$, here $S$ is the cross section area of the wire. 
The DW is not stable against a rigid shift to the right because this small change in spin 
structure always lower the system energy. Thus a DW must vary with time under a magnetic field.

When $J$ is much larger than the Zeeman energy, $\boldsymbol{M}_1$ and $\boldsymbol{M}_2$ 
are always anti-parallel to each other. We define $\boldsymbol{M}_{\mathrm{eff}}=(M_1-M_2) 
\boldsymbol{m}$, where $\boldsymbol{m}$ is the unit vector of $\boldsymbol{M}_1$. 
Then $\boldsymbol{m}$ satisfies the following equation
 \begin{equation}\label{torque}
(s_1 - s_2) \frac{\partial \boldsymbol{m} }{\partial t} =- (M_1 -M_2) 
 \boldsymbol{m} \times \boldsymbol{H}_{\mathrm{eff}} + \alpha 
 \boldsymbol{m} \times \frac{\partial \boldsymbol{m} }{\partial t}, 
\end{equation}
where $\boldsymbol{H}_{\mathrm{eff}}=\frac{M_1 \boldsymbol{H}_1-M_2\boldsymbol{H}_2}
{M_1 -M_2}$. In terms of $\boldsymbol{m}$, the total energy is $E[\boldsymbol{m}]=\int 
\left[A(\nabla \boldsymbol{m})^2+f(\boldsymbol{m})-\mu_0(M_1 -M_2)\boldsymbol{m} 
\cdot \boldsymbol{H} \right] d^3 \boldsymbol{x}$ with $A =A_1 
M_1^2 +A_2 M_2^2$, where $a$ is the lattice constant. 
Denote $\alpha = \alpha_{11} s_1+\alpha_{22}s_2-\alpha_{12}s_2 \frac{\gamma_{2}}
{\gamma_{1}} - \alpha_{21} s_{1}\frac{\gamma_{1}}{\gamma_{2}}$, the thermodynamic 
second law requires $\alpha>0$ to ensure the Rayleigh dissipation functional 
$\mathcal{R}=\frac{\mu_0\alpha}{2}\int\left( \frac{\partial\boldsymbol{m}}{\partial t} 
\right)^2 d^3 \boldsymbol{x}$ \cite{Yuan1,AFM1,gilbert} to be positive-definite.
Equation \eqref{torque} says that the change of spin angular momentum (left-hand side) 
equals the net torque (right-hand side) that is the sum of a torque from an effective 
field on the net magnetization ($M_1 -M_2\neq 0$) and a dissipative torque from the 
motion of $\boldsymbol{m}$. At the AMCP, the dissipative torque cancels the field torque. 

Equation \eqref{torque} can be recast as an effective LLG
equation \cite{effLLG1, effLLG2, effLLG3, effLLG4,exper3} 
\begin{equation} \label{effLLG}
 \frac{\partial \boldsymbol{m}}{\partial t} = - \gamma_{\mathrm{eff}} \boldsymbol{m} \times 
\boldsymbol{H}_{\mathrm{eff}} + \alpha_{\mathrm{eff}} 
 \boldsymbol{m} \times \frac{\partial \boldsymbol{m}}{\partial t},
\end{equation}
with an effective gyromagnetic ratio $\gamma_{\mathrm{eff}}=|M_1 - M_2|/(s_1 - s_2)$ 
and an effective Gilbert damping $\alpha_{\mathrm{eff}}=\alpha /(s_1 - s_2)$. 
$\gamma_{\mathrm{eff}} \alpha_{\mathrm{eff}}$ is always positive because a moving 
magnetization must dissipate its energy to its environment (See Eq. \eqref{dEdt} below).
$s_1 > s_2 $ and $s_1 < s_2 $ correspond to lattice-1 and lattice-2 dominate cases.  
Following a similar derivation in the literature \cite{ZZS1,ZZS2}, the energy 
dissipation rate is \cite{xrw1,xrw2},
\begin{equation} \label{dEdt}
\frac{dE}{dt}=-\frac{\alpha_{\mathrm{eff}} \gamma_{\mathrm{eff}} \mu_0  }{\left(1+ 
\alpha_{\mathrm{eff}}^2 \right) (M_1 - M_2)} \int \left(\boldsymbol{M}_{\mathrm{eff}}
 \times \boldsymbol{H}_{\mathrm{eff}} \right)^{2} d^3 \boldsymbol{x}.
\end{equation}

We divide the wire into three regions as shown in Fig. \ref{fig1}: I for the domain with 
$\boldsymbol{M}_{\mathrm{eff}}$ parallel to $\boldsymbol{H}$, II for the DW, and III 
for the domain with $\boldsymbol{M}_{\mathrm{eff}}$ anti-parallel to $\boldsymbol{H}$.
Energy dissipation occurs only in the DW region (region II) where $\boldsymbol{M}_
{\mathrm{eff}}$ and $\boldsymbol{H}_{\mathrm{eff}}$ are not collinear \cite{xrw1,xrw2}.
The change of energies, $E_{\mathrm{I}}$ and $E_{\mathrm{III}}$, of region I and III 
comes from the DW propagation along the wire, and should be $\frac{d(E_{\mathrm{I}} +
E_{\mathrm{III}})}{dt}=-2 \mu_0 (M_1 -M_2 ) H v S$, where $v$ is the DW velocity. 
DW energy $E_{\mathrm{II}}$ must be around a certain value. Thus the time averaged 
energy change rate must be zero. In another word, $\frac{d E_{\mathrm{II}}}{dt}$ is 
either zero or oscillates with zero average. The energy conservation requires   
\begin{equation}\label{v1}
\begin{aligned}
v=&\frac{\alpha_{\mathrm{eff}} \gamma_{\mathrm{eff}} }{2H S \left(1+ \alpha_{\mathrm{eff}}^2 \right)} 
\int \left(\boldsymbol{m} \times \boldsymbol{H}_{\mathrm{eff}} \right)^{2} d^3 \boldsymbol{x} \\& 
+\frac{1}{2 \mu_0 (M_1 -M_2 ) H  S}\frac{dE_{\mathrm{II}}}{dt}.
\end{aligned}
\end{equation}
This is a universal relationship between DW velocity and the DW structure and  can 
serve as a proper definition of instantaneous DW velocity.
The second term on the right should be identical zero in the case of a rigid DW 
motion such that the DW speed is constant. In the case that a DW deforms itself 
during its propagation, the energy dissipation rate and DW energy $E_{\mathrm{II}}$ 
oscillate with time and $\overline{\frac{dE_{\mathrm{II}}}{dt}}=0$, where the bar 
denotes the time average. This results in an oscillating DW speed whose 
time-averaged value is 
 \begin{equation}\label{vbar}
\bar{v}=\frac{\alpha_{\mathrm{eff}} \gamma_{\mathrm{eff}} }
{2H S \left(1+ \alpha_{\mathrm{eff}}^2 \right)} 
 \overline{ \int \left(\boldsymbol{m} \times \boldsymbol{H}_{\mathrm{eff}}
  \right)^{2}  d^3 \boldsymbol{x}  }.
\end{equation}

We note $\left(\boldsymbol{m}\times\boldsymbol{H}_{\mathrm{eff}} \right)^2=H_{\mathrm{eff},
\theta}^2+H_{\mathrm{eff},\phi}^2$, where $H_{\mathrm{eff}, \theta}$ and $H_{\mathrm{eff},
\phi}$ are two field components perpendicular to $\boldsymbol{m}$ in the local 
coordinate framework $\left(\boldsymbol{e}_{\boldsymbol{m}}, \boldsymbol{e}_\theta,
\boldsymbol{e}_\phi\right)$. $\theta(\boldsymbol{x},t)$ and $\phi(\boldsymbol{x},t)$ are 
the polar and the azimuthal angles of $\boldsymbol{m}$. 
Below we consider anisotropy energy of $f(\boldsymbol{m})=-K_z m_z^2 +K_y m_y^2 $. 
\begin{equation}
\begin{aligned}\label{Hthetaphi}
H_{\mathrm{eff},\theta}&=H \sin \theta-G, \\
H_{\mathrm{eff},\phi}&= -\frac{1}{\mu_{0} (M_1 - M_2) \sin \theta} \frac{\partial f}
{\partial \phi} \\& \quad +\frac{2  A}{\mu_{0} (M_1 - M_2) \sin \theta}\frac{\partial}
{\partial z} \left(\sin ^{2}\theta \frac{\partial \phi}{\partial z} \right),
\end{aligned} 
\end{equation}
where $G=\frac{1}{\mu_0(M_1-M_2)}\left[ 2A\frac{\partial^2\theta}{\partial z^2}-\frac{\partial f}
{\partial\theta}-2A\sin\theta\cos\theta\left( \frac{\partial\phi}{\partial z}\right)^{2}\right]$.

Equation \eqref{effLLG} along $\boldsymbol{e}_{\theta},\boldsymbol{e}_{\phi}$ becomes 
\begin{equation}
\begin{aligned}\label{LLG2}
\frac{\partial \theta}{\partial t} & =\gamma_{\mathrm{eff}} H_{\mathrm{eff},\phi}-\alpha_{\mathrm{eff}}
\sin \theta \frac{\partial \phi}{\partial t}\\
\sin \theta \frac{\partial \phi}{\partial t}& =-\gamma_{\mathrm{eff}} H_{\mathrm{eff},\theta}
+\alpha_{\mathrm{eff}} \frac{\partial \theta}{\partial t}.
\end{aligned}
\end{equation}
Eliminate time-derivative of $\theta$ from equation \eqref{LLG2}, we have  
\begin{equation}\label{eq15}
\left(1+\alpha_{\mathrm{eff}}^2 \right) \sin \theta 
\frac{\partial \phi }{\partial t}=\gamma_{\mathrm{eff}} \left(\alpha_{\mathrm{eff}} 
H_{\mathrm{eff},\phi}-H_{\mathrm{eff},\theta} \right) . 
\end{equation}
If the DW propagates as a rigid-body along the $z$-direction, the case of a field 
below the Walker breakdown \cite{Walker}, i.e. $\frac{\partial \phi}{\partial z}=0$, 
$\frac{\partial^2 \phi}{\partial z^2}=0$, and $\frac{\partial \phi}{\partial t}=0$, 
using Eq. \eqref{Hthetaphi}, we have $2A\frac{\partial^2 \theta}{\partial z^2}- 
\frac{\partial f}{\partial \theta} =0$, so that $H_{\mathrm{eff},\theta}=-H 
\sin\theta$, whose maximal allowed external field is the Walker breakdown field. 
For $(M_1 -M_2), (s_1 -s_2)\neq 0$, $\frac{\partial \phi}{\partial t}=0$ obviously 
requires $ \alpha_{\mathrm{eff}} H_{\mathrm{eff},\phi}=H_{\mathrm{eff},\theta}$. 
This means that the DW-plane cants an angle to generate a non-zero $H_{\mathrm{eff},
\phi}$ to coherently vary $\theta$ such that the DW propagates along the wire.  
Recall our biaxial model $f(\theta,\phi)=-K_z\cos^2 \theta+K_y\sin^2\theta\sin^2 \phi$, 
the Walker breakdown field is $H_W=\max\left(\frac{\alpha_{\mathrm{eff}}K_y\sin 2\phi}
{\mu_0 (M_1 - M_2 )} \right)=\frac{\alpha  K_y}{\mu_0 |(M_1 - M_2 )(s_1 -s_2)|}$. 
Both $\gamma_{\mathrm{eff}}$ and $\alpha_{\mathrm{eff}}$ diverge as $(s_1 -s_2)^{-1}$ near 
the AMCP. The limit of Eq. \eqref{eq15} under $s_1-s_2\to 0$ ($\alpha_{\mathrm{eff}}^2,
\gamma_{\mathrm{eff}}\alpha_{\mathrm{eff}} \sim(s_1-s_2)^{-2}$) gives $H_{\mathrm{eff},
\phi}=0$ and $\phi=0$ when $\frac{\partial \phi}{\partial t}=0$. Thus the DW-plane 
remains in the $xz$-plane and never rotates, leading to an infinite $H_W$ at the AMCP.  
One can also see this point by considering an equivalent form of Eq. \eqref{effLLG},
$\frac{\partial \boldsymbol{m}}{\partial t} =-\frac{\gamma_{\mathrm{eff}}}
{1+\alpha_{\mathrm{eff}}^2}\boldsymbol{m}\times \boldsymbol{H}_{\mathrm{eff}}
-\frac{\gamma_{\mathrm{eff}}\alpha_{\mathrm{eff}}}{1+\alpha_{\mathrm{eff}}^2}
\boldsymbol{m}\times \left(\boldsymbol{m}\times \boldsymbol{H}_{\mathrm{eff}}\right)$.
At the AMCP, the precessional torque vanishes since $\frac{\gamma_{\mathrm{eff}}}
{1+\alpha_{\mathrm{eff}}^2} \to 0$ as $(s_1 -s_2)\to 0$ while the damping torque 
is finite because of $\lim_{(s_1 -s_2 ) \to 0} \frac{ \gamma_{\mathrm{eff}}
\alpha_{\mathrm{eff}}}{1+\alpha_{\mathrm{eff}}^2}=\frac{M_1 - M_2}{\alpha}\neq 0$. 
This means that the precessional motion is completely prohibited, and $\boldsymbol{m}$ 
at any point inside the DW rotates coherently toward external field, leading to a 
rigid DW propagation along the wire (see the Supplemental Materials \cite{supp}).

Equation \eqref{effLLG} with our biaxial magnetic anisotropy has the well-known Walker 
DW solution \cite{Walker} of $\theta(z,t)= 2\arctan \left(\exp \left( \left(z-\int_{0}
^{t}v(\tau) d\tau \right)/\Delta(t)  \right)\right)$, where $\Delta$ is the DW width.  
It gives $\frac{dE_{\mathrm{II}}}{dt}=- \frac{4  A S}{\Delta^2} \frac{d \Delta}{dt}=0$ 
for a rigid-body DW propagation. $H_{\mathrm{eff},\theta}=\alpha_{\mathrm{eff}} 
H_{\mathrm{eff},\phi}=-H\sin \theta$, $\left(\boldsymbol{m} \times \boldsymbol{H}_
{\mathrm{eff}} \right)^{2} =H^{2}_{\mathrm{eff},\theta}+H^{2}_{\mathrm{eff},\phi}=
\left( 1+\alpha_{\mathrm{eff}}^2 \right)H^2 \sin^2  \theta / \alpha_{\mathrm{eff}}^2 $. 
Substituting DW width definition of $\int \sin^2 \theta d^3 \boldsymbol{x} =2 S \Delta$  
into Eq. \eqref{v1}, one has $v=\frac{(M_1 -M_2 ) \Delta}{\alpha}H$ and DW speed 
at Walker breakdown field $v_{W}=\frac{K_y \Delta }{\mu_0 (s_1-s_2)}$, independent of 
the damping coefficient and divergent at the AMCP.

Away from the AMCP, $H_W$ is finite. A DW shall precess around wire axis while it 
propagates along the wire when $H>H_W$. From Eqs. \eqref{vbar} and \eqref{Hthetaphi}, 
we have 
\begin{equation}\label{eqexpandv}
\bar{v}=\frac{\alpha_{\mathrm{eff}} \gamma_{\mathrm{eff}}}
{2H S \left(1+ \alpha_{\mathrm{eff}}^2 \right)}\overline{\int
\left[(H\sin\theta -G)^2 +H_{\mathrm{eff},\phi}^2 \right] d^3 \boldsymbol{x}}. 
\end{equation}
Average DW velocity is (see the Supplemental Materials \cite{supp} for detailed derivation), 
\begin{equation}\label{v-formula}
\bar{v}=c_1 H + \frac{c_1}
 {\alpha_{\mathrm{eff}}^2}\left( H- \sqrt{H^2 -H_W^2} \right)
\end{equation}
where $c_1=\frac{\alpha_{\mathrm{eff}} \gamma_{\mathrm{eff}} }{2 S \left(1+ 
\alpha_{\mathrm{eff}}^2 \right)} \overline{\int \sin^2 \theta d^3 \boldsymbol{x}}=
\frac{(M_1-M_2) {\alpha}\bar{\Delta}}{(s_1-s_2)^2+ {\alpha}^2}$ is peaked at the AMCP. 
Equation \eqref{v-formula} is exact under very sensible assumptions, and all coefficients 
in Eq. \eqref{v-formula} are fully determined by the model parameters. 

\begin{table*}[htbp]
\setlength{\tabcolsep}{6mm}{
\begin{tabular}{ccccccc}
\hline\hline\noalign{\smallskip}
Data set  & Set 1 & Set 2 & Set 3 & Set 4 & Set 5 & Set 6
\\ \noalign{\smallskip}\hline\noalign{\smallskip}
$K_{y,1}  (\,\mathrm{MJ}/\mathrm{m}^3) $   &    0.05      &  0.035  &   0.02 & 0.1 & 0.1 & 0.1         \\ 
$K_{y,2}  (\,\mathrm{MJ}/\mathrm{m}^3) $    &    0.05      &  0.035  &   0.02 & 0.1 & 0.1 & 0.1         \\ 
$\alpha_{11}$        & 0.02    & 0.02     & 0.02    &    0.005 &         0.01 &        0.015            \\
$\alpha_{22}$        & 0.02    & 0.02     & 0.02    &    0.005 &         0.01 &        0.015            \\
$\alpha_{\mathrm{eff}}$  &     0.3473      &  0.3473   &  0.3473   & 0.0868 & 0.1736 & 0.2605       \\
$K_{y}  (\,\mathrm{MJ}/\mathrm{m}^3) $    &    0.1      &  0.07  &   0.04 & 0.2 & 0.2 & 0.2        \\ 
$\mu_0 H_W$ (T)   &     0.3157     & 0.2210   &  0.1263  & 0.1579 & 0.3157 & 0.4736         \\ 
$\bar{\Delta}$(nm)   &     3.85      &  3.87   &  3.89   &  3.79  &  3.75  &  3.79     \\ 
$c_1 (\mu_0 \cdot \mathrm{m}\cdot \mathrm{s}^{-1} 
\mathrm{T}^{-1}) $  &     210.00      &  211.13   &  212.34  &  57.48 &  111.37 &    162.69    \\
\noalign{\smallskip}\hline
\end{tabular} }
\caption{$K_{y,1}$, $K_{y,2}$, $\alpha_{11}$, and $\alpha_{22}$ are model parameters. 
$\alpha_{\mathrm{eff}}$, $K_{y}$, $\mu_0 H_W$, $\bar{\Delta}$, and $c_1$ are computed quantities. }
\label{table1}
\end{table*}

Equation \eqref{v-formula} predicts a negative differential DW mobility in the range of 
$H_W<H< \frac{\alpha^2_{\mathrm{eff}} +1}{\sqrt{ \alpha^4_{\mathrm{eff}}+2
\alpha^2_{\mathrm{eff}} }}H_W$. This prediction is also true for 
ferromagnetic case. In order to find out how accurate of Eq. \eqref{v-formula} is for 
$H>H_W$, we use MuMax3 \cite{mumax3} to numerically solve Eq. \eqref{dynamics} for a 
synthetic ferrimagnetic strip wire as shown in Fig. \ref{fig1} that consist of two 
antiferromagnetically-coupled ferromagnetic-layers of 1nm thick each. 
The strip size is $16 \mathrm{\, nm}\times 2\mathrm{\, nm}\times 1024\mathrm{\, nm}$. 
The cell size in simulations is chosen to be $1\mathrm{\, nm}\times 1\mathrm{\, nm}\times 
1\mathrm{\, nm}$. To mimic a GdFeCo alloy \cite{exper1}, the model parameters are $J=1.2 
\times 10^{-4}\, \mathrm{J}\cdot\mathrm{A}^{-2}\mathrm{m}^{-1}$, $A_1=9.8\times 10^{-24} 
\mathrm{\, J}\cdot\mathrm{m}\cdot\mathrm{A}^{-2}$, $A_2=1.23\times 10^{-23}\mathrm{\, J}
\cdot \mathrm{m} \cdot \mathrm{A}^{-2}$, biaxial anisotropy are considered for each 
sublattice, $f_i = -\frac{K_{z,i}}{M_i^2}M^2_{i,z}+\frac{K_{y,i}}{M_i^2}M^2_{i,y}$,
$i=1,2$, $K_{z,1}=K_{z,2}=0.65 \,\mathrm{MJ}/\mathrm{m}^3$, $\alpha_{12}=\alpha_{21}=0$.  
$K_{y,i}$ and $\alpha_{ii}$ ($i=1,2$) are used for simulating different systems as 
labelled by Set 1-6 in Table \ref{table1}. The gyromagnetic ratios $\gamma_1=\gamma_2
=1.76 \times 10^{11} \, \mathrm{s}^{-1} \mathrm{T}^{-1}$, the saturation magnetizations 
are $ M_1 =1010 \,\mathrm{kA}/\mathrm{m}$, $ M_2 =900 \,\mathrm{kA}/\mathrm{m}$.
The coupling field between two sublattices is of hundreds of Tesla to guarantee 
collinearity of two spin sublattices. Different from a natural ferimagnet, 
inter-sublattice coupling is along the $y$-direction in our synthetic ferrimagnet.
In the simulation, a DW is first created at the center of nanowire, then a uniform
magnetic field is applied in the $+\hat{z}$ direction. The velocity is obtained 
from the linear fit of time-evolution curve of the DW center (where $m_z =0$).
For high fields above Walker breakdown, the average velocities are obtained from 
data accumulated for more than 4 velocity oscillating periods. 

\begin{figure}
\centering
\includegraphics[width=8.5cm]{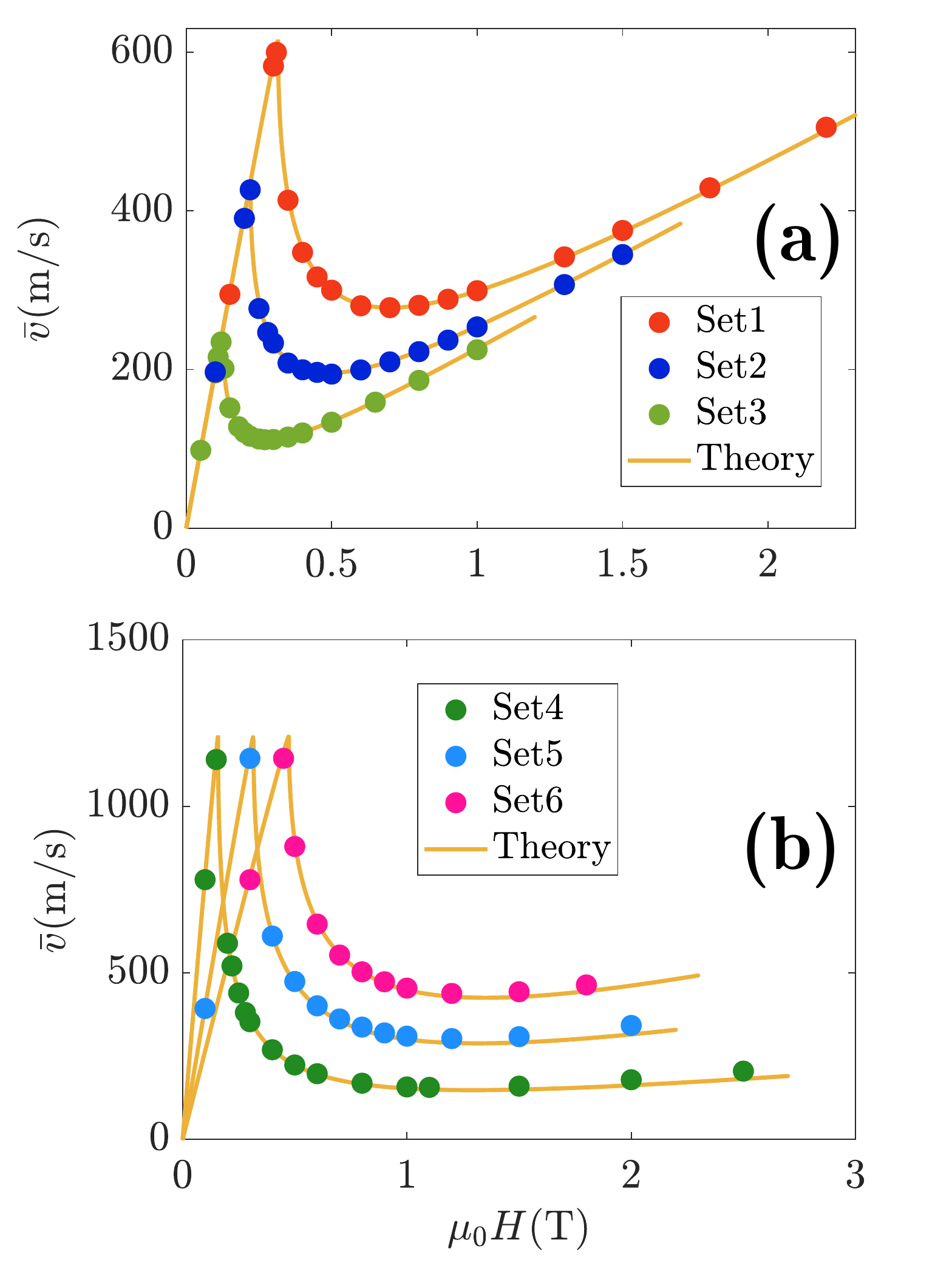}\\
\caption{Average DW speed of a head-to-head DW as a function of an applied field 
along the $+\hat{z}$-direction. Symbols are MuMax3 simulation results and solid curves 
are theoretical formula without any fitting parameter. 
a) Three systems (denoted as Set 1, 2, 3) with the 
same $\alpha_{11}=\alpha_{22}=0.02$ but different $K_{y,i}$.
b) Three systems with (denoted as Set 4, 5, 6) with the same $K_{y,i}=
0.1\mathrm{\,MJ}/\mathrm{m}^3$ but different $\alpha_{ii}$ ($i=1,2$). 
Their values are listed in Table \ref{table1}.}
\label{fig2}
\end{figure}

We consider six different systems with various $K_{y,i}$ and $\alpha_{ii}$ ($i=1,2$).
The detail values of the model parameters are given in Table \ref{table1}.
Because of large speed difference, Fig. \ref{fig2}(a) plot $\bar v$ vs. $H$ for 
three systems with the same $\alpha_{ii}=0.02$ and different $K_{y,i}$, label as Set 
1, 2, 3. Figure \ref{fig2}(b) is the similar plots for three systems with the same 
$K_{y,i}=0.1\mathrm{\,MJ}/\mathrm{m}^3$, but different $\alpha_{ii}$, label as Set 
4, 5, 6. The corresponding values of $c_1$, $\alpha_{\mathrm{eff}}$, and $H_W$ 
computed from this theory are also given in Table \ref{table1}. The perfect 
agreements between the simulation results (the symbols) and theoretical prediction 
(the solid curves) demonstrate that Eq. \eqref{v-formula} is almost exact. 

\textit{Discussion and Conclusion}.---
Before conclusion, we would like to make a few remarks. 1) The relationship 
between the instantaneous DW speed and the DW structure is exact that explains   
why our high-field DW speed formula without any fitting parameters agree 
perfectly with simulation results.
2) Since no collective-mode approximation is used, the theory is applicable to  
all types of DWs.   
3) High DW speed is a result of the absence of Walker breakdown field at the AMCP. 
This explains the observed high DW speed of more than 1.5km/s at the AMCP although 
the mobility $\mu=\frac{(M_1-M_2)\Delta}{\alpha}$ for $H<H_W$ itself is comparable 
to or even smaller than that for a ferromagnetic wire \cite{FMDW1,FMDW2}.  

In summary, a generic theory of field-driven DW motion in FiM wires is presented.
A static DW cannot exist in a homogeneous ferrimagnetic nanowire when 
a uniform static magnetic field or any other external force creates an 
energy density difference between two domains separated by the DW. 
Spins in the DW must vary with time under the external magnetic field 
such that the system energy is dissipated due to the Gilbert damping. 
The dissipated energy must be compensated by the Zeeman energy released 
from moving the DW toward the domain with the higher energy density. 
High DW speed near the AMCP is the consequence of the absences of  
precessional torque and infinite high Walker breakdown field at the AMCP. 
A lower Zeeman energy density and a high energy dissipation rate contribute 
also to the high DW speed at a reasonable lower field near the AMCP. 
Away from the AMCP, our approach can not only obtain the exact DW velocity 
below the Walker breakdown field, but also an almost exact velocity formula 
beyond the Walker breakdown field. This theory agrees with all existing 
experiments and simulations, and provides useful guidances to DW manipulation.  

\textit{Acknowledgments}.---
This work is supported by the National Key Research and Development Program of 
China 2020YFA0309600, the National Natural Science Foundation of China (Grant 
No. 11974296) and Hong Kong RGC Grants (No. 16301518, 16301619 and 6302321).

\end{document}